\documentclass[twocolumn]{revtex4}
\usepackage{amsfonts,color}
\usepackage{graphicx,amsmath,amssymb}
\def\T{\mathbf{T}}
\begin{document}
\title{Experimental pre--assessing entanglement in Gaussian states mixing}
\author{Adriana~Pecoraro}
\email{adriana.pecoraro@fisica.unina.it}
\affiliation{Dipartimento di Scienze Fisiche, 
Universit\`{a} \textquotedblleft Federico II\textquotedblright , 
Complesso Univ. Monte Sant'Angelo, I-80126 Napoli, Italy}
\author{Daniela~Buono}
\email{danielabuono@yahoo.com}
\affiliation{Dipartimento di Ingegneria 
Industriale, Universit\`a degli Studi di Salerno, 
I-84084 Fisciano (SA), Italy}
\author{Gaetano~Nocerino}
\email{gaetanocerino@libero.it}
\affiliation{Trenitalia spa, DPR Campania, 
IMC Campi Flegrei, I-80124 Napoli, Italy}
\author{Alberto~Porzio}
\email{alberto.porzio@spin.cnr.it}
\affiliation{Dipartimento di Scienze Fisiche, 
Universit\`{a} \textquotedblleft Federico II\textquotedblright , 
Complesso Univ. Monte Sant'Angelo, I-80126 Napoli, Italy}
\affiliation{
CNR -- SPIN, Unit\'{a} di Napoli, Complesso Univ. Monte Sant'Angelo,
I-80126 Napoli, Italy}
\author{Stefano~Olivares}
\email{stefano.olivares@mi.infn.it}
\affiliation{Quantum Technology Lab, 
Dipartimento di Fisica dell'Universit\`a 
degli Studi di Milano, I-20133 Milano, Italia.}
\affiliation{Istituto Nazionale di Fisica Nucleare, Sezione di Milano,
I-20133 Milano, Italia.}
\author{Matteo~G.~A.~Paris}
\email{matteo.paris@fisica.unimi.it}
\affiliation{Quantum Technology Lab, 
Dipartimento di Fisica dell'Universit\`a 
degli Studi di Milano, I-20133 Milano, Italia.}
\affiliation{Istituto Nazionale di Fisica Nucleare, Sezione di Milano,
I-20133 Milano, Italia.}
\date{\today}
\begin{abstract}
We suggest and demonstrate a method to assess entanglement generation
schemes based on mixing of Gaussian states at a beam splitter (BS). Our
method is based on the fidelity criterion and represents a tool to
analyze the effect of losses and noise before the BS in both symmetric
and asymmetric channels with and without thermal effects.  More
generally, our scheme allows one to pre-assess  entanglement resources
and to optimize the design of BS-based schemes for the generation of
continuous variable entanglement.
\end{abstract}
\maketitle
\section{Introduction}
Continuous-variable (CV) entanglement is a powerful resource for optical
quantum technologies and it is the crucial ingredient for several
protocols including quantum teleportation, dense coding and
quantum-enhanced metrology. In particular, Gaussian states and Gaussian
entanglement revealed themselves as the main resource in practical
applications \cite{Brau,rev12,FOP05,g2,g3,g4,g5,Wolf,Olivares2012}.  In
fact, most of CV quantum technology has been developed upon exploiting
Gaussian states and Gaussian operations. In this framework, techniques
for the generation, the characterization and the certification of
Gaussian entanglement play a crucial role and received a large attention
in the last two decades.
\par
Among the different schemes to generate CV entanglement 
Gaussian states emerged as a convenient choice, and it has been employed
in several applications. In this scheme, a pair of Gaussian states is
mixed at a beam splitter and entangled states are obtained at the output
as far as the input signals show some nonclassical features
\cite{nclee0,nclee,nclee1,kim02,Wxb02,Furusawa98,Masada15}.
Entanglement certification is usually performed {\em a posteriori}, by
means of measurements realized at the outputs of the BS, e.g. by
performing full quantum tomography of the bipartite state
\cite{por09,cia16} or by measuring a suitable entanglement witness.  On
the other hand, an {\em a priori} certification scheme, i.e. involving
measurements before the BS, would be welcome since it would permit the
pre-assessing of entanglement resources and the optimization of the
generation scheme, e.g. by quantum state or reservoir engineering at the 
input.
\par
In this paper, we experimentally address an {\em a priori} certification
scheme based on the fidelity criterion for entanglement generation
\cite{Olivares2011}. In particular, we assess the effects of losses and
noise on the generation of entanglement by Gaussian states mixing. Upon
using a suitably designed experimental setup, we analyze the effect of
signals propagation before the BS and evaluate threshold values on the
transmission coefficient and on the thermal noise as a function of the
parameter of the input signals. We consider both symmetric and
asymmetric channels with and without thermal noise.
\par
The paper is structured as follows. In Section \ref{s:mix} we introduce
the notation and the tools to describe entanglement generation by
Gaussian states mixing.  We also review the fidelity criterion proposed
in \cite{Olivares2011}.  In Section \ref{s:prop} we describe Gaussian
states propagation in lossy and noisy channels.  In Section \ref{s:exp}
we describe our experimental apparatus and report results for the
different configurations. Section \ref{s:out} closes the paper with some
concluding remarks.
\section{The fidelity criterion} 
\label{s:mix}
As a matter of fact, there are several \textit{a posteriori} 
criteria to witness entanglement of bipartite CV systems 
\cite{PHS,PHS1,PHS2,Duan2000,Reid1989}. They may be exploited
to assess whether entanglement has been produced \textit{after} 
a given interaction took place. On the other hand,
an \textit{a priori} criterion has been recently introduced
\cite{Olivares2011} to assess the entanglement capability of 
a pair of distinct Gaussian states interacting at a BS.
The criterion is based on the mutual fidelity of two 
uncorrelated  Gaussian states, $\varrho_c$ and
$\varrho_d$, describing the preparation of 
two bosonic modes $c$ and $d$, and is able to predict 
whether the state obtained by
mixing them at a BS, will be entangled or not.  More in detail, this
criterion states that the interaction between the two input states
through a bilinear exchange Hamiltonian, gives rise to entanglement iff
the fidelity between the two input states is less than a threshold value
$F_{th}$.  The actual value for the threshold depends on the
initial states purities and on the beam splitter transmissivity 
\cite{Olivares2011}.
\par
In order to review the criterion, let us first review some properties
of Gaussian states (GS). GS are quantum states with a Gaussian Wigner
function in the phase space.  Gaussian states are prominent 
resources in CV quantum technology because, besides being 
easily produced in laboratories with current technologies, 
they preserve their Gaussian character under linear and 
bilinear transformations, such as those associated with 
beam splitters, phase shifters and optical amplifiers, e.g. 
single- and two-mode squeezing \cite{Olivares12}, as well 
as during propagation through a noisy channel  \cite{PRA2012}.
Gaussian states are completely characterized by a finite 
number of parameters, e.g. by the first and second moments 
of quadrature mode operators. In particular, when dealing 
with the correlation between the two modes of a
bipartite Gaussian state, one may focus only on the 
phase-space CM, since the presence of first moments does 
not affect the amount of correlations.
\par
Let us now consider 
the phase-space description of the two
involved modes, and of their dynamics. As we already mentioned, we can
focus the evolution of the $4 \times 4$ CM  $\Sigma_{\rm in}$ of the the
input states $\varrho_c \otimes \varrho_d$ (without loss of generality
we can focus on states with vanishing first-moment values of the the
quadrature operators). If $\sigma_k$ refers to the CM of the mode
$k=c,d$, then $\Sigma_{\rm in} = \sigma_c \oplus \sigma_d$. After the
evolution, characterized by the parameter $\tau \in [0,1]$ (the BS
transmissivity), the evolved CM can
be written in the following block-matrix form 
\cite{Olivares2011,Olivares2012}
\begin{align}
\Sigma_{\rm out}=\left(\begin{array}{cc}
\Sigma_{1} & \Sigma_{12}\\
\Sigma_{12}^T & \Sigma_{2}
\end{array}\right)
\end{align}
whose elements are:
\begin{subequations}\label{evolved:CM:BS}
\begin{align}
&\Sigma_{1}=\tau\sigma_{c}+(1-\tau)\sigma_{d},\\
&\Sigma_{2}=\tau\sigma_{d}+(1-\tau)\sigma_{c},\\
&\Sigma_{12}= \tau(1-\tau)(\sigma_{d}-\sigma_{c}).
\end{align}
\end{subequations}
The presence of the non-zero off-diagonal terms $\Sigma_{12}$, suggests
the presence of correlations between the two output modes, whose amount 
depends on how much the incoming single mode matrix are different.
It is worth noting that if $\sigma_c = \sigma_d$, then the evolved
states is uncorrelated: this effect has been theoretically
\cite{Olivares09,Olivares12} and experimentally investigated for
different kind of Gaussian states \cite{Bloomer11,Meda13}. Therefore,
the correlations arise if and only if $\sigma_c \ne \sigma_d$. This can
be translated into a more quantitative expression by introducing the
fidelity of the two initial Gaussian states written for their covariance
matrices \cite{Scutaru98}:
\begin{align}
F_{cd}= \frac{1}{\sqrt{\Delta+\delta}-\sqrt{\delta}}
\end{align}
where 
\begin{align}
\Delta&=\det[\sigma_{c}+\sigma_{d}], \\
\delta&=\left\{\det[\sigma_{c}]-\frac{1}{4}\right\}
\left\{\det[\sigma_{d}]-\frac{1}{4}\right\}\,.
\end{align}
In Ref.~\cite{Olivares2011} it has been proved that iff this quantity
falls under the following threshold:
\begin{align}
F_{th} = \frac{4\mu_c \mu_d \sqrt{\tau(1-\tau)}}
{\sqrt{g_{-}+4 \tau(1-\tau) g_{+}}-\sqrt{4 \tau(1-\tau) g_{-}}},
\end{align}
where 
$$g_{\pm}\equiv g_{\pm}(\mu_c,\mu_d)=\prod_{k=c,d}(1\pm\mu_k^2)\,,$$
$\mu_k=$, $k=c,d$ being the purities of the two local states, 
then entanglement is generated, and the bipartite system 
emerging from the mixing is not separable. 
The inequality
\begin{equation}
F_{cd} \leq F_{th}\,,
\end{equation}
thus represents a necessary and sufficient criterion to 
pre-assess entanglement resources, i.e. to assess them
before they are actually employed in generation of entanglement
by mixing at a beam splitter.
The threshold depends on the input states purities $\mu_k$ and on 
the BS transmissivity $\tau$.
\section{Propagation in noisy channels} \label{s:prop}
As a matter of fact, pure quantum states cannot be effectively produced
in a laboratory. The unavoidable technical imperfections and the
interaction with the environment induce decoherence on any state, which
evolves into a statistical mixture, even if pure at the time of its
generation. Pure quantum features, such as entanglement, are strongly
affected by this mechanism, and therefore it is of fundamental interest
to know how the interaction with the environment alters the parameters
determining the \textit{quantumness} of a state. 
\par
In quantum optical systems, loss of photons is a relevant source of
decoherence. In addition, one may model the environment as a thermal
bath made of infinite modes at thermal equilibrium and, in the most
general case, each containing some residual squeezing
\cite{kennedy88,tombesi94,lutkenhaus98,rossi04}.  In the context of an
open systems approach \cite{Breur2002}, upon assuming weak coupling with
the environment and the absence of any memory effect, the evolution of a
single-mode propagating through a Gaussian noisy transmission channel is
well described by a Master Equation (ME) in the Lindblad form. The 
ME may be then rewritten in terms of a Fokker-Planck (FP) equation for 
the Wigner quasi-probability distribution \cite{FOP05}. Since we are
dealing with (zero mean) Gaussian states the full information about
the evolved state is contained in the time evolution of the CM, 
which reads 
\begin{align}
\sigma(t)=
\sqrt{\mathbb{G}_t} \, \sigma(0)
\sqrt{\mathbb{G}_t} 
+(\mathbb{I}-\mathbb{G}_t)
\sigma_{\infty}\label{eq:evoluition}\,,
\end{align}
where $\sigma(0)$ is the initial CM, 
$\mathbb{G}_t = e^{-\Gamma_{c}t}\mathbb{I}  
\oplus e^{-\Gamma_{d}t}\mathbb{I}$, 
$\Gamma_k$ being the damping rate of mode $k=c,d$, 
and $\sigma_{\infty}=\sigma_{c,\infty} \oplus\sigma_{d,\infty}$ with:
\begin{align}
\label{asympt:CM:k}
\sigma_{k,\infty}=\left(
\begin{array}{cc}
(\frac{1}{2}+N_{k})+\text{Re}[M_k] & \text{Im}[M_{k}]\\
\text{Im}[M_{k}] & (\frac{1}{2}+N_{k})-\text{Re}[M_{k}]
\end{array}
\right)
\end{align}
represents the diffusion matrix and the asymptotic CM of the system,
i.e. the CM for $t\rightarrow\infty$.  In Eq.~(\ref{asympt:CM:k})
$N_{k}$ and $M_{k}$ represent the effective photons number and the
squeezing parameter of the bath interacting with mode $k$, respectively.
It is worth noting that, to ensure the positivity of the density matrix
associated with the evolved state, one should have $| M_{k} |^2 \le
N_{k} (1+N_{k})$.
Equation (\ref{eq:evoluition}) suggests that the
action of the lossy channel with damping rate on the CM characterizing
each single system is in all equivalent to the action of a fictitious BS
that couples each mode $k=c,d$ to the corresponding environment through its
transmission coefficient $\T_k=e^{-\Gamma_k t}$.
This simple picture does not depend on the property of the bath.
\section{Experiment and results}
\label{s:exp}
The a priori entanglement criterion discussed in Section 2 has been
experimentally tested for a balanced BS, i.e. for $\tau=1/2$ in Eqs.
(\ref{evolved:CM:BS}). In particular, we have analysed the mixing
of pairs of squeezed modes which are subject 
to different transmission channels.
The two initial fields are obtained by optically manipulating the output
of a frequency degenerate type-II optical parametric oscillator (OPO) 
working below threshold.
The experimental setup  \cite{APB2008} is based on a continuous-wave (cw)
Nd:YAG laser  (Innolight-Diabolo dual wavelength) internally frequency
doubled.  The second harmonic ($532$ nm) is employed as OPO pump. The
non-linear crystal of the OPO is a $1 \times 1.5 \times 25$ mm$^3$
periodically poled $\alpha$-cut KTP crystal (PPKTP) manufactured by
Raicol Crystals Ltd. on custom design. The use of the $\alpha$-cut
PPKTP allows implementing a type II phase matching with cross-polarized
signal and idler waves. The frequency-degeneracy condition, $\lambda_i =
\lambda_s = 2 \lambda_p = 1064$ nm (IR), is achieved at $ T \sim 326$ K.
The crystal temperature is actively controlled, while 
a Pound-Drever-Hall system \cite{PDH} controls the OPO length in 
order to ensure pump-cavity resonance. 
\par
Our device generates an entangled bipartite Gaussian state consisting of
two collinear beams (signal and idler) corresponding to two orthogonally
polarized modes $a$ and $b$, each excited in a thermal state
\cite{IJQI2007}. Since the Hamiltonian underlying the process is $H
\propto (a^\dag b^\dag + \text{h.c.})$, that is the Hamiltonian of
two-mode squeezing, by suitably manipulate the polarization of the modes
$a$ and $b$ and let them interfere as described in \cite{JOB2005}, it is
possible to obtain two independent squeezed fields, corresponding to the
states of the modes $$c=(a+b)/\sqrt{2} \qquad d=(a-b)/\sqrt{2}\,.$$
In particular, the modes $c$ and $d$ show squeezing in orthogonal phase
quadrature and, thus, represent a pair of single-mode squeezed states
able to generate entanglement in a mixing process, in a scheme already
exploited in the first implementation of the CV teleportation protocol
\cite{Furusawa98}.  Since modes $c$ and $d$ are, in general,
\emph{uncorrelated} squeezed states,
characterized by the single-mode CMs:
\begin{subequations}\label{eq:cm}
\begin{align}
\sigma_{c} &= (2\mu_{c})^{-1} \, \text{diag}\left[e^{2r},e^{-2r}\right], \\
\sigma_{d} &= (2\mu_{d})^{-1} \, \text{diag}\left[e^{-2r},e^{2r}\right],
\end{align}
\end{subequations}
respectively, $r$ being the so-called squeezing parameter and $\mu_k$
the purity of mode $k=c,d$, they can be used to test the fidelity
criterion.  However, aiming at a more general overview of the above
mentioned criterion, we have considered different scenarios that have
been investigated both theoretically and experimentally.
\par
Experimental tests have been conducted by considering different sets of
squeezed thermal states at the output of the type--II OPO (see
Refs.~\cite{PRA2012,JOSAB2010} for details).  In particular, we have
selected pairs of squeezed states satisfying different conditions on
their squeezing parameter and/or thermal contribution, depending on the
different investigated scenarios.
\begin{figure}[h!]
\begin{center}
\includegraphics[width=0.97\columnwidth]{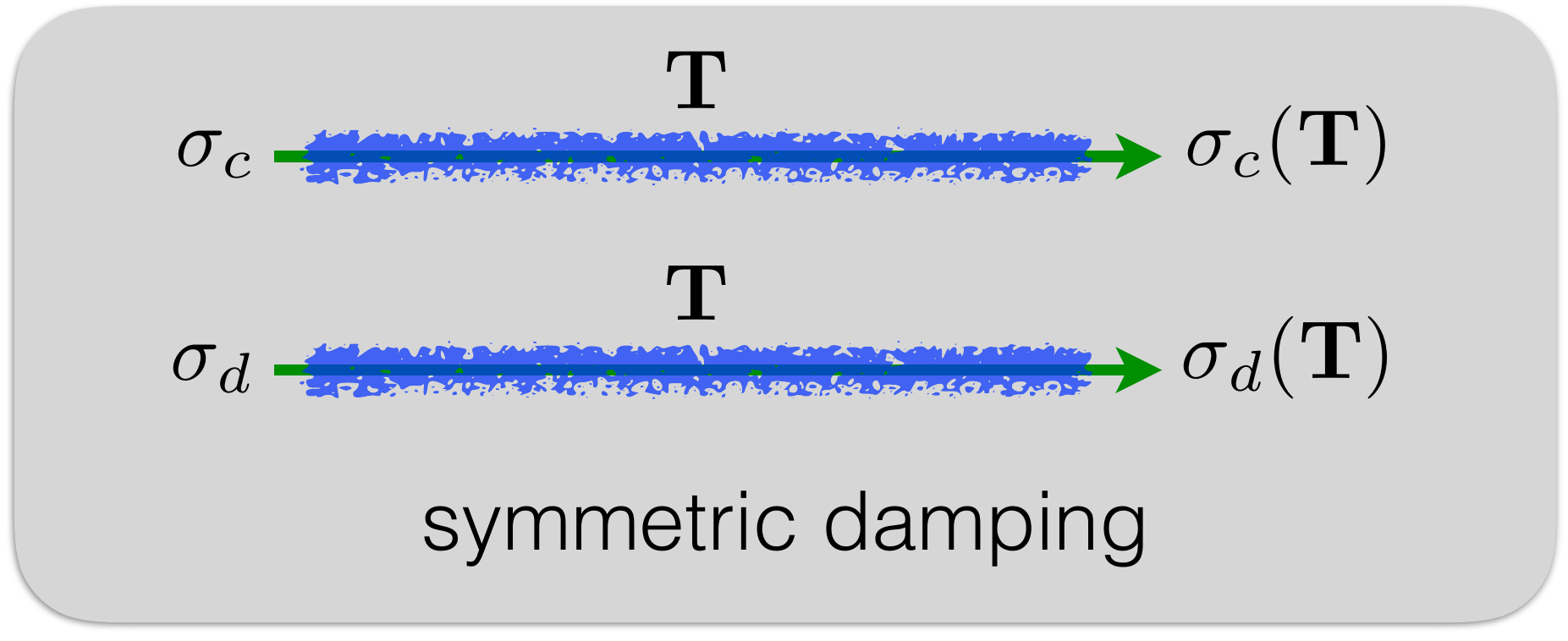}
\end{center}
\caption{Schematic diagram of the symmetric damping channel. Modes $c$
and $d$ propagate through identical passive channels of transmissivity
$\mathbf{T}$.} \label{Fig:symmetric_scheme}
\end{figure}
\subsection{Symmetric Passive Damping}
\begin{figure}[h!]
\begin{center}
\includegraphics[width=0.97\columnwidth]{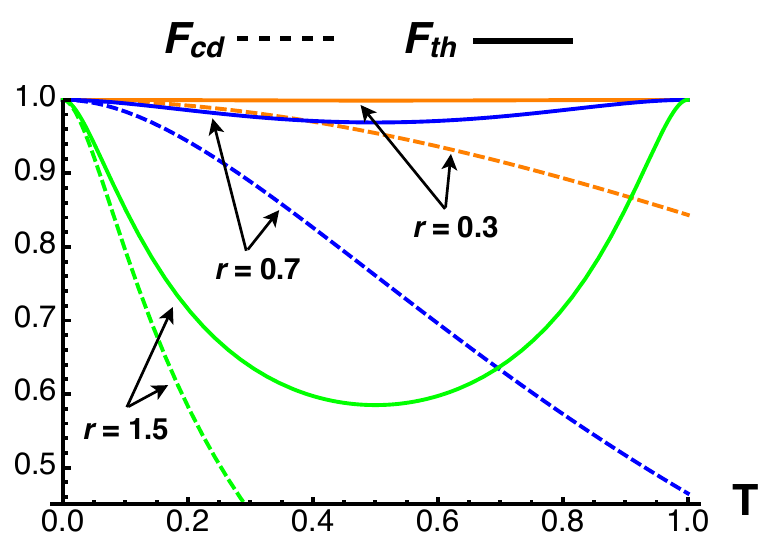}
\end{center}
\caption{Plot of the theoretical behavior of $F_{cd}$ (dashed
lines) and the corresponding threshold condition $F_{th}$ (solid lines)
at the output of two identical transmission channels as functions
of transmission $\mathbf{T}$ of the channels, for three
different values of initial squeezing $r$ of the initial pure
states ($\mu_{c} = \mu_{d} = 1$).}
\label{Fig:symmetric_theory}
\end{figure}

The simplest scenario is a symmetric damping. The two uncorrelated modes
undergo to the same passive damping (see
Fig.\ref{Fig:symmetric_scheme}).  In such a case, the evolution of both
the threshold and the fidelity among the two states depends on the
squeezing parameter of the ancestor pure squeezed state \cite{LP2014}.
In Fig.~\ref{Fig:symmetric_theory} we show the expected behavior of the
threshold fidelities (solid lines) and of the actual fidelities (dashed
lines) as function of $\mathbf{T}$ (the channel transmission) for
different value of the initial squeezing parameter $r$.  Theoretical
behaviors have been obtained by considering the single mode CM
evolution of Eq.
(\ref{eq:evoluition}) that, in our case, reads (for the sake of
simplicity we drop the subscripts $c$ and $d$):
\begin{align}
\sigma(\mathbf{T})=\mathbf{T}\sigma(0) +
(1-\mathbf{T})\frac{1}{2}\mathbb{I}\label{eq:evoluzione} 
\end{align}
being $\mathbf{T}=e^{-\Gamma t}$ the transmission coefficient and
$\frac{1}{2}\mathbb{I}$ the vacuum state CM.
\par
Figure~\ref{Fig:symmetric_theory} shows that fidelities approach to 1 when the two modes are
maximally attenuated ($\mathbf{T} \to 0$): in this case, as clear from expression (\ref{eq:evoluzione}),
they become two vacuum states; in addition, we can see that the
maximum of difference between the two fidelities occurs in correspondence
of fully transmitted pure states ($\mathbf{T} \to 1$).
Moreover, we can see that attenuation alone is not enough to prevent
initially pure squeezed states to give rise to an entangled pair,
since $F_{th} > F_{cd}$, $\forall\, \mathbf{T} \in (0,1]$ \cite{PRA2012}.
As we will see later, only by setting the system in contact with a thermal
reservoir we can observe the violation of the entanglement condition.
\par
\begin{figure}[h!]
\begin{center}
\includegraphics[width=0.97\columnwidth]{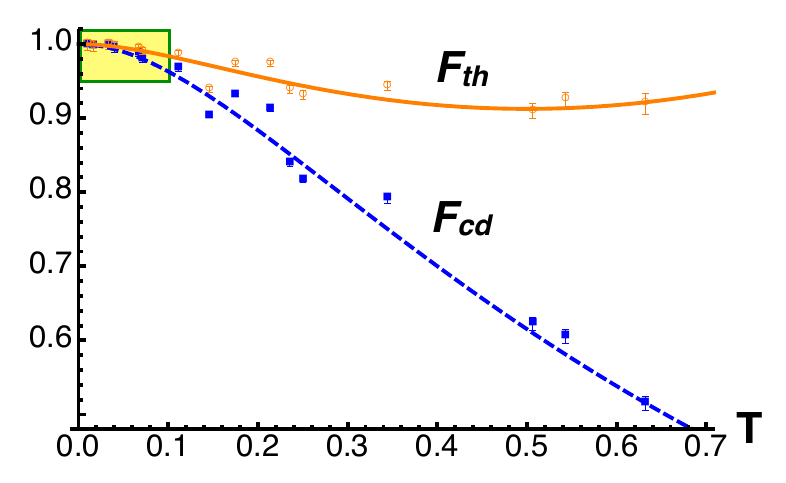}
\includegraphics[width=0.97\columnwidth]{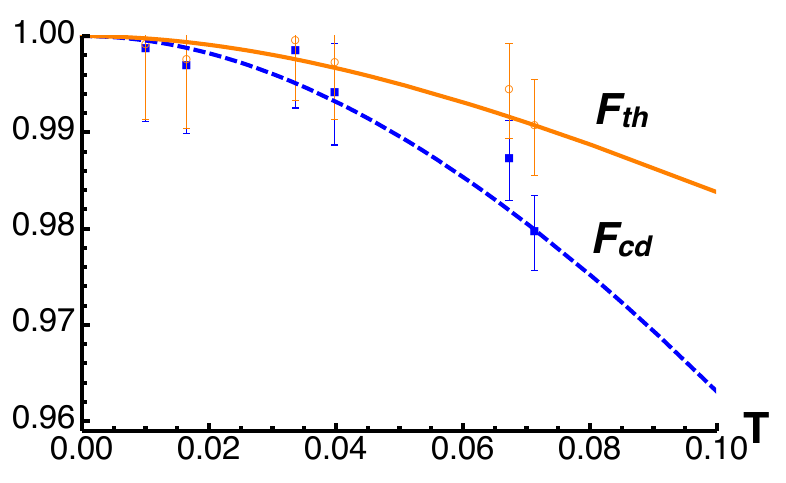}
\end{center}
\caption{Upper panel: Plot of the theoretical $F_{cd}$ (dashed line),
threshold condition $F_{th}$ (solid line) and corresponding experimental
data (symbols) as functions of $\mathbf{T}$ when the two modes $c$ and
$d$ pass through two identical transmission channels (see
Fig.~\protect\ref{Fig:symmetric_scheme}).  Experimental data refer to pairs of
states showing the same value for the squeezing parameter ($r \simeq
0.92$) of the initial pure state. Lower panel: Magnification of the
region highlighted in the upper left corner of the upper panel.}
\label{Fig:symmetric_exp}
\end{figure}

In order to experimentally assess the {\it a priori} fidelity criterion,
we have evaluated the fidelities for pairs of modes with orthogonal
squeezing phases, after they had experienced the same passive damping,
the noisy channel being simulated by a variable attenuator (mimicking
the BS) and the characterization obtained by a homodyne detector
\cite{pscrp}.
In Fig.~\ref{Fig:symmetric_exp} (upper panel) we report the experimental
behavior obtained for two modes undergoing the same attenuation.
Experimental states have been {selected} from a larger set by
calculating for each state the initial squeezing parameter $r$
\cite{JPB2006} and the actual value for the transmission $\mathbf{T}$.
Then, pairs of states having the same value
of $r$ within the experimental uncertainties, have been used for the plot.
Contrarily to other CV separability criteria, the one investigated here
depends on the states themselves. So, also the behavior of the
threshold value has been experimentally verified.  
\par
In the lower panel Fig. 
of \ref{Fig:symmetric_exp} we show a zoom of the high absorption region
($\mathbf{T} <0.1$) to show that, by applying a posteriori criteria, CV
entanglement can be set between very low energy states, as also verified
in \cite{pscrp}.  The plots show a good agreement between the reported
data and the theoretical expectation. Experimental uncertainties 
have been evaluated by means of the usual propagation formula.
According to the theorem, if noise only springs from the photons loss,
mixing two orthogonally squeezed modes in a BS always gives rise to
entangled bipartite states exploitable for quantum communication tasks
independently from
squeezing level \cite{kim02,Wxb02}. 
\par
\begin{figure}[h!]
\begin{center}
\includegraphics[width=0.97\columnwidth]{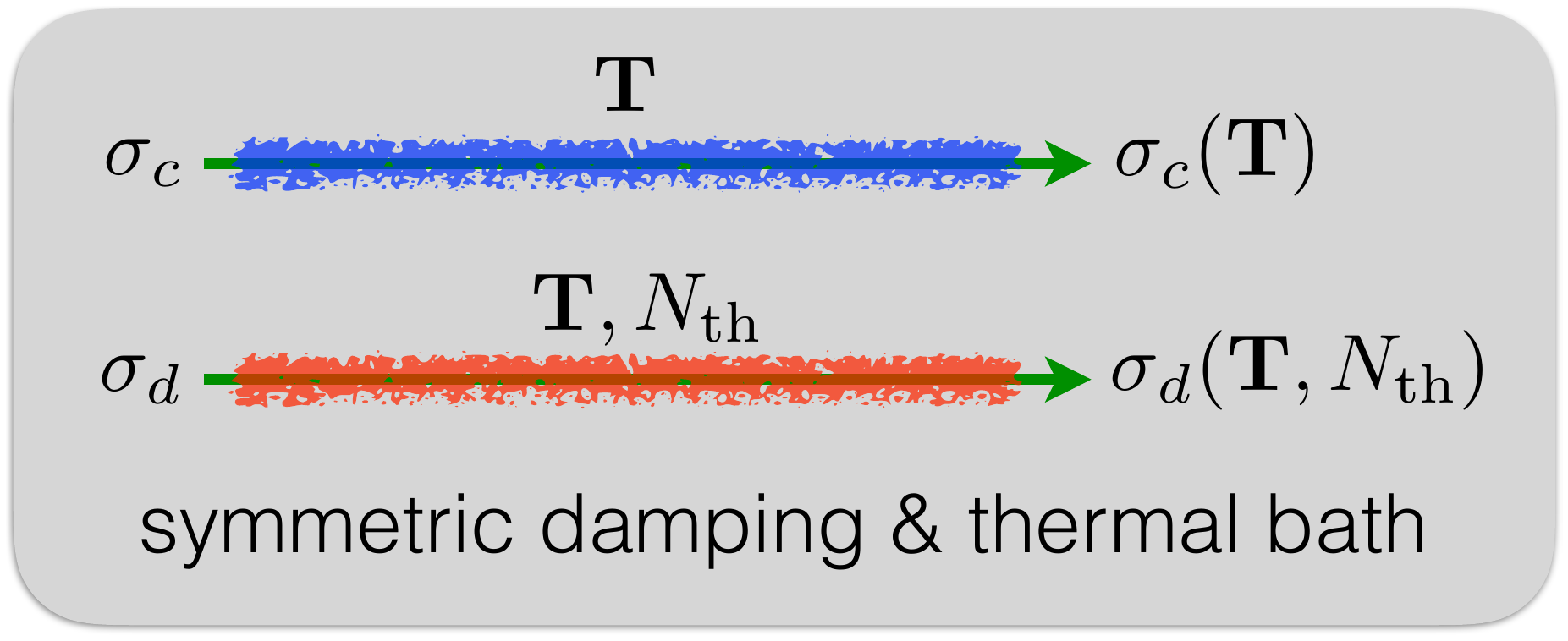}
\end{center}
\caption{Schematic diagram of the symmetric damping channel 
with thermal
noise.  Modes $c$ and $d$ propagate through identical passive channels
of transmissivity  $\mathbf{T}$. One of the mode is also coupled to a
thermal bath with $N_{\rm th}=1$ average thermal photons.}
\label{Fig:thermal_scheme}
\end{figure}
\subsection{Symmetric damping with thermal noise}
Here we consider the case where one couples one of the two modes to a
thermal bath with non-zero mean photon number, i.e. characterized by a
non-zero effective temperature.  As sketched in
Fig.~\ref{Fig:thermal_scheme}, mode $c$ travels a Gaussian transmission
channel devoid of thermal noise; so its evolution is described by Eq.
(\ref{eq:evoluzione}). Differently mode $d$ besides attenuation is
coupled to a thermal bath with a given average number of thermal photons
$N_{\rm th}$. To model its evolution we have to replace the vacuum CM in
Eq. (\ref{eq:evoluzione}) by the CM corresponding to a thermal state
$\sigma_{\infty} = (\frac{1}{2}+N_{\rm th}) \mathbb{1}_2$, so that:
\begin{align}
\sigma_{d}(\mathbf{T})=\T\sigma_{d}(0)+(1-\T)\sigma_{\infty}\, .
\end{align}
We stress that, in this case, the lower is the transmission the higher
is the number of thermal photons that couples into the mode from the
unused port of the BS.
\par
\begin{figure}[tb!]
\begin{center}
\includegraphics[width=0.97\columnwidth]{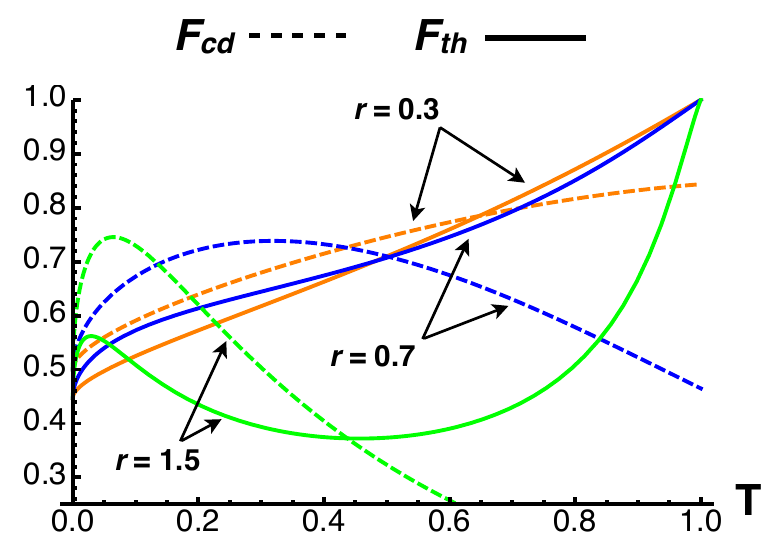}
\end{center}
\caption{Plots of $F_{cd}$ and $F_{th}$
(dashed and solid lines) as functions of $\mathbf{T}$ for different
values of initial squeezing $r$ in presence of a thermal bath with
$N_{\rm th}=1.0$ coupled to only one of the
two modes.}
\label{Fig:thermal_theory}
\end{figure}

The introduction of a thermal bath changes dramatically the scenario.
As shown in Fig.~\ref{Fig:thermal_theory}, the channel may now be
entanglement breaking \cite{EB:giovannetti} and, given the initial
squeezing parameter, we find a limiting value for the transmissivity.
For lower transmissivity the thermal photons coupled to one of the mode
prevent the birth of entanglement (see Fig.~\ref{Fig:thermal_theory}).
Indeed, for each of the three
squeezing parameters we have selected, we can identify a well defined
region where entanglement would not be attainable by mixing the two modes:
the higher is the squeezing parameter the smaller is such a region.
\par
\begin{figure}[tb!]
\begin{center}
\includegraphics[width=0.97\columnwidth]{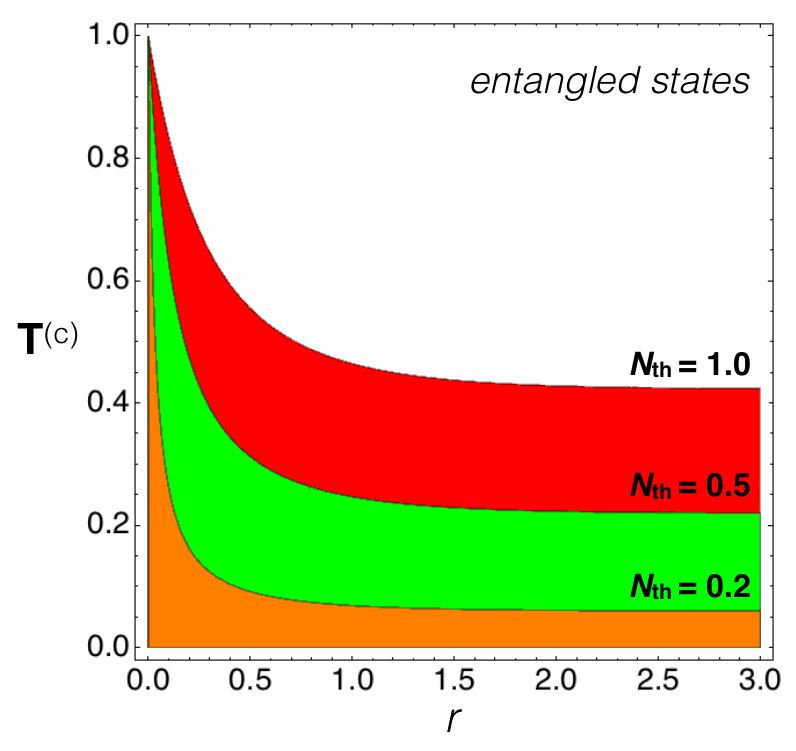}
\end{center}
\caption{Critical transmission $\mathbf{T}^{\rm (c)}$ as a function of $r$
for different values of  $N_{\rm th}$ (from top to bottom $N_{\rm th} = 1.0$, $0.5$ and $0.2$).
The colored regions below the curves show where
entanglement cannot be achieved, whereas above 
the curves one obtains
always  an entangled output state.}
\label{Fig:thermal_intesection_Tvsr}
\end{figure}
In Fig.~\ref{Fig:thermal_intesection_Tvsr}
we report the value for the critical transmission, i.e. the value
$\mathbf{T}^{\rm (c)}$ at which happens intersection between the
fidelity and the threshold as a function of $r$ the initial squeezing
parameter for a thermal bath with three different values of $N_{\rm
th}$.  The curves split the parameter space into two region: above the
curve entanglement may be obtained whereas in the lower colored regions 
it cannot.  As
one may expect, the value of the transmission coefficient, associated
with the transition, is lower for higher $r$, so the more the modes are
squeezed the larger is the interval in which they are good as
entanglement resources. We also note that in the limit of large
squeezing ($r \gg1$), the critical transmission tends to the following
asymptotic values depending only on the number of thermal photons of the
bath $N_{\rm th}$:
\begin{align}
\mathbf{T}^{\rm (c)} \to 1 - \frac{\sqrt{1 + N_{\rm th}(2+9 N_{\rm th})}-1 + N_{\rm th}}{2N_{\rm th}(1+2 N_{\rm th})}\quad (r \gg1).
\end{align}
This sheds some light on the fundamental role of squeezing in producing entanglement
by mixing two independent modes. As we have seen (Fig.~\ref{Fig:symmetric_theory}),
in the case of two pure states ($\mathbf{T}=1$) we had $F_{\rm th}=1$,
thus any pair of even slightly different pure Gaussian
states is a good entanglement resource.
On the contrary the presence of thermal noise requires that at least one of the two modes is squeezed by some amount in order to produce entanglement.
\par
\begin{figure}[h!]
\begin{center}
\includegraphics[width=0.97\columnwidth]{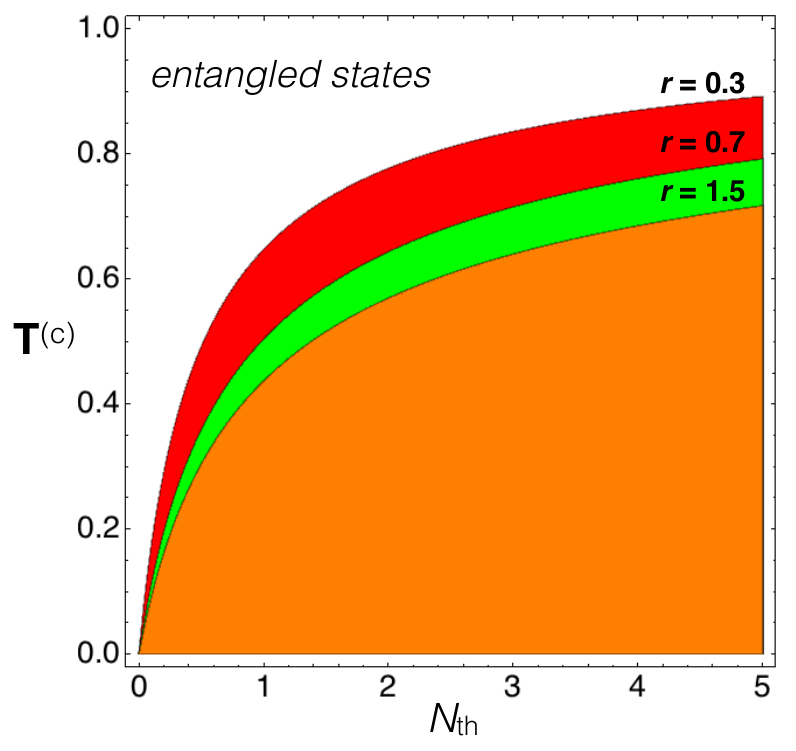}
\end{center}
\caption{Critical transmission $\mathbf{T}^{\rm (c)}$ as a function of
$N_{\rm th}$, the average thermal photon number of the bath, for
different values of the squeezing parameter, from to to bottom $r=0.5$,
$0.7$ and $1.5$.  The curves divide the parameter plane into two
regions: above 
the lines entanglement can be achieved, below (colored regions) it cannot. }
\label{Fig:thermal_intesection_TvsNth}
\end{figure}

The same critical transmission has been evaluated for a given squeezing
parameter and a variable number of thermal photons of the bath (see
Fig.~\ref{Fig:thermal_intesection_TvsNth}).  The increase of the thermal
photons would reduce the possibility of obtaining entanglement in mixing
the two modes.
\subsection{Asymmetric damping}
\begin{figure}[tb!]
\begin{center}
\includegraphics[width=0.97\columnwidth]{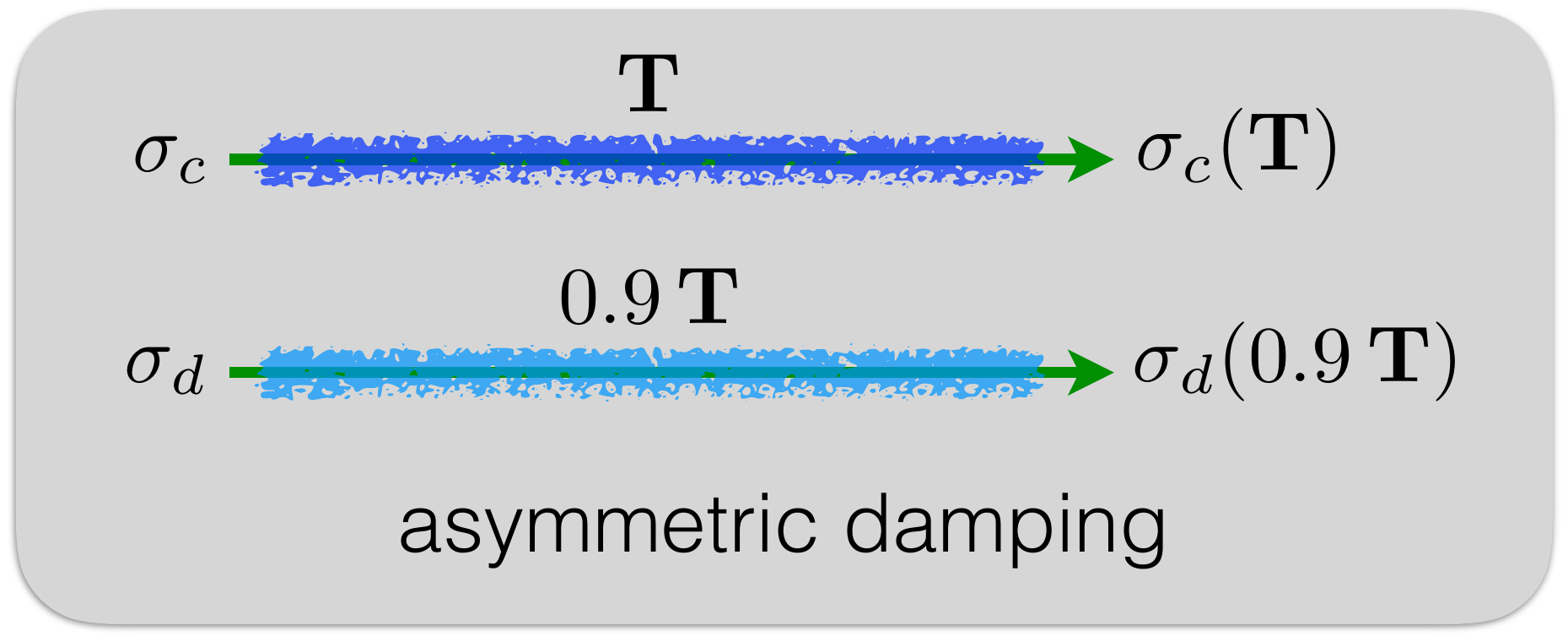}
\end{center}
\caption{
Schematic diagram of the asymmetric damping channel. 	Modes $c$ and
$d$ propagate through two channels of different transmissivity. In
particular, for any value of $\mathbf{T}$ we assume that mode $c$ is
transmitted through a channel with transmission $\mathbf{T}$, whereas
mode $d$ propagates through
a channel with transmission $0.9\, \mathbf{T}$.}
\label{Fig:asymmetric_scheme}
\end{figure}

In entanglement distribution schemes \cite{ed1,ed2,ed3,ed4} it may
happen that the channels have different transmissivities. In our case,
this means that the two squeezed fields arrive at the mixing BS after
having suffered different attenuations. This situation is sketched in
Fig.~\ref{Fig:asymmetric_scheme} where we assume that mode $d$ travels
along a channel with a transmission $0.9 \mathbf{T}$, if 
$\mathbf{T}$ is the value corresponding to mode $c$.
\par
\begin{figure}[h!]
\begin{center}
\includegraphics[width=0.97\columnwidth]{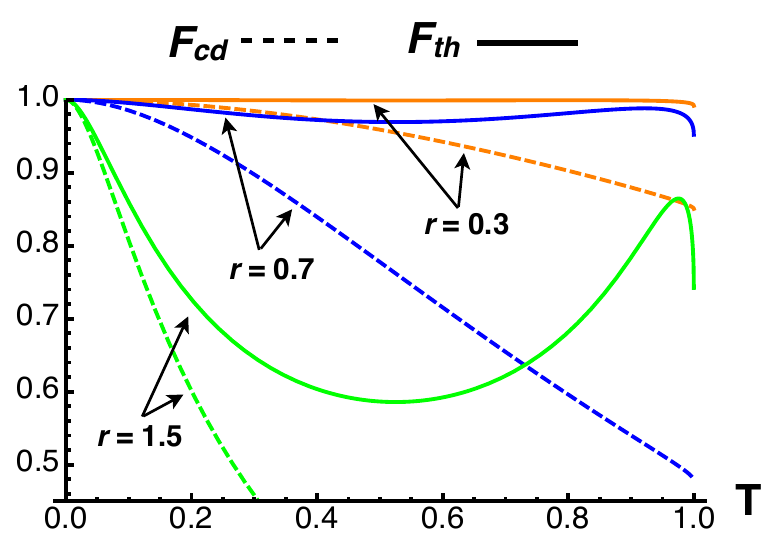}
\end{center}
\caption{Plot of the theoretical behavior of $F_{cd}$ and $F_{th}$
(dashed and solid lines, respectively) as functions of $\mathbf{T}$ for
different values of the initial squeezing $r$. Modes $c$ and $d$
propagate through different channel with transmissivities $\mathbf{T}$
and $0.9 \mathbf{T}$, respectively.}
\label{Fig:asymmetric_theory}
\end{figure}

The effects of such an asymmetry can be seen in
Fig.~\ref{Fig:asymmetric_theory}.  It is evident that the transmission
asymmetry does not play any effective role in corrupting the two states
properties.  Compared to the symmetric case discussed in
Fig.~\ref{Fig:symmetric_theory}, the only clear difference can be found
in the threshold fidelity $F_{\rm th}$ for modes affected by lower
losses ($\mathbf{T} \gtrsim 0.9$): mode $c$ purity is close to $1$ while
mode $d$ has already suffered and effective decoherence.
\par
\begin{figure}[h!]
\begin{center}
\includegraphics[width=0.97\columnwidth]{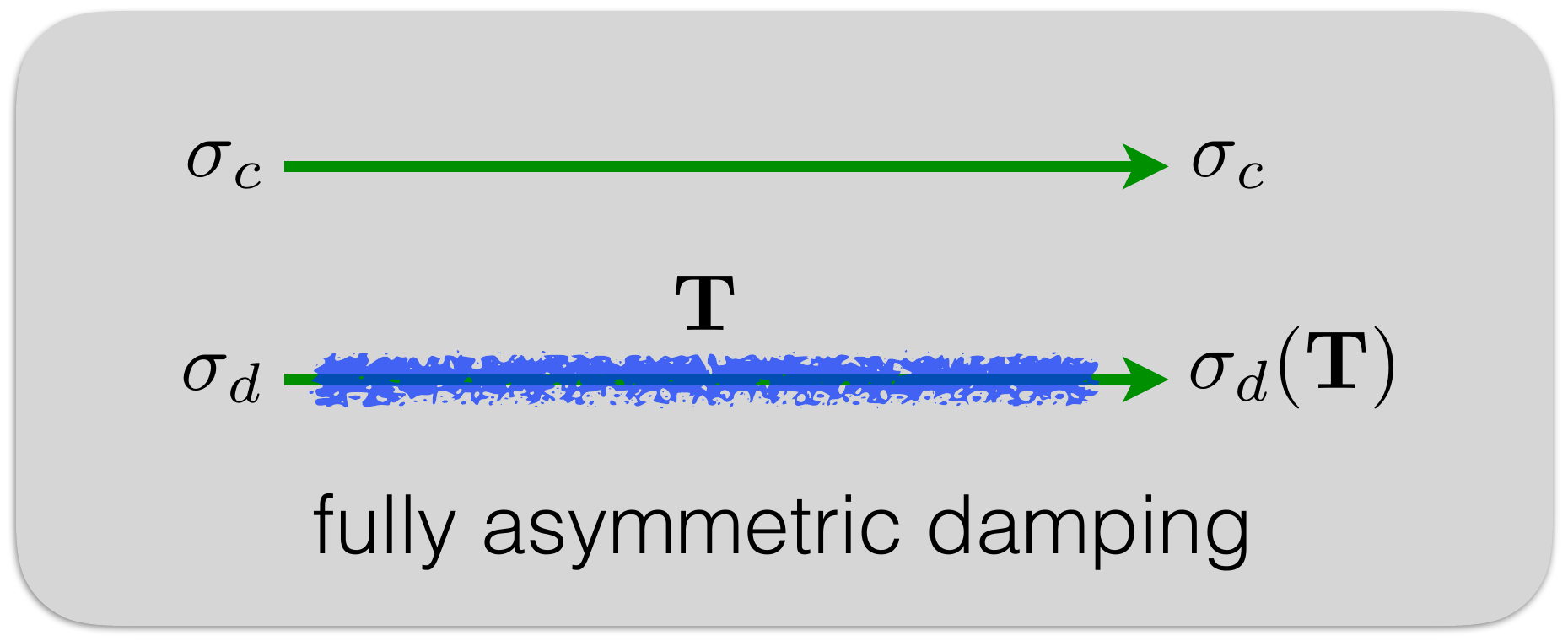}
\end{center}
\caption{
Schematic diagram of the fully asymmetric damping channel. Mode $c$
propagates in an ideal lossless channel (and stays pure), whereas mode
$d$ travels a passive transmission channel with transmission coefficient
$\mathbf{T}$.} \label{Fig:asymmetric_scheme2}
\end{figure}
\begin{figure}[h!]
\begin{center}
\includegraphics[width=0.97\columnwidth]{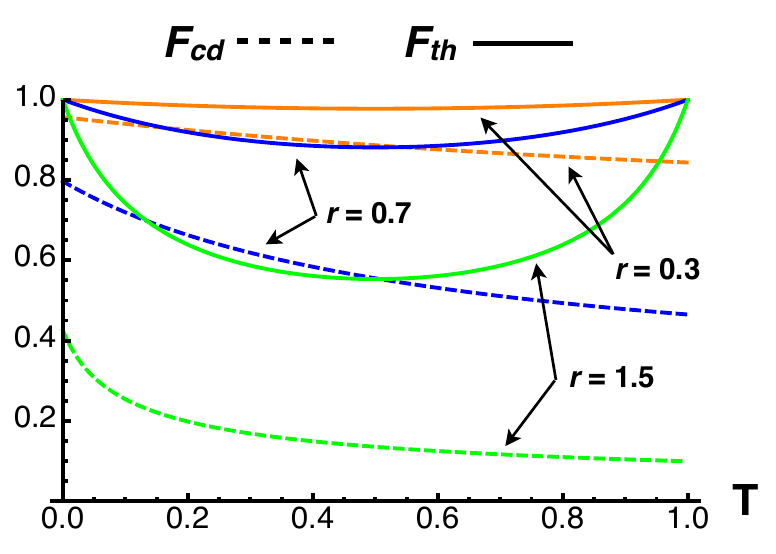}
\end{center}
	\caption{Plot of the theoretical behavior of $F_{cd}$ (dashed
	line) and $F_{th}$ (solid line) as functions of $\mathbf{T}$ and
	different values of the initial squeezing $r$ for the fully
	asymmetric damping scheme of Fig.~\protect
	\ref{Fig:asymmetric_scheme2}.}
	\label{Fig:asymmetric_theory2} \end{figure}
\begin{figure}[tb!]
\begin{center}
\includegraphics[width=0.97\columnwidth]{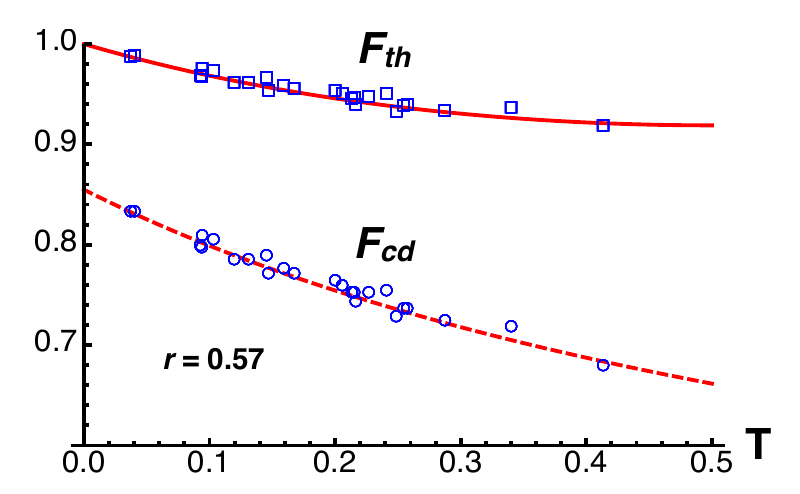}
\end{center}
	\caption{Plot of the theoretical (lines) and experimental
	(squares and circles) fidelity $F_{cd}$ (dashed line) and the
	threshold condition (solid line) as functions of $\mathbf{T}$,
	when one of the two modes remains pure while the other one
	travels a passive transmission channel (see
	Fig.~\protect\ref{Fig:asymmetric_scheme2}).  Experimental data refer to
	a set of measured CMs showing the same value of the squeezing
	parameter ($r= 0.57$) for the ancestor pure state.}
	\label{Fig:asymmetric_experimental2}
\end{figure}

This effect can be more clearly understood considering a scenario in
which one of the two modes stays pure while the other one experiences a
passive channel of transmission $\mathbf{T}$, as sketched in
Fig.~\ref{Fig:asymmetric_scheme2}).  The results, shown in
Fig.~\ref{Fig:asymmetric_theory2}, prove once more that the two modes
outing a balanced BS illuminated by a single squeezed field (with the
vacuum entering through its unused port) are entangled. As it can be
seen in Fig.~\ref{Fig:asymmetric_theory2} for any value of initial
squeezing at $\mathbf{T} = 0$ (namely, mode $d$ is in the vacuum state)
we have $F_{cd} < F_{th}$.
This scenario has been also investigated experimentally. We have
addressed the fidelity among a single pure ancestor state, retrieved
from the experiment, and the whole set of the mixed states corresponding
to the same pure ancestor but for different effective channel
transmissions.  The experimental curve has been obtained selecting, among
the measured CMs, all those corresponding to the transmission of 
a pure squeezed state with $r = 0.57$ (within
experimental uncertainty).  The corresponding plot is reported,
together with the corresponding theoretical curve, in
Fig.~\ref{Fig:asymmetric_experimental2}: even at very low transmission
the measured fidelity is well below the threshold value, revealing 
the generation of entanglement at the output of a balanced 
BS illuminated by a single mode squeezed field.
\par
\subsection{Asymmetric damping with thermal noise}
\begin{figure}[h!]
\begin{center}
\includegraphics[width=0.97\columnwidth]{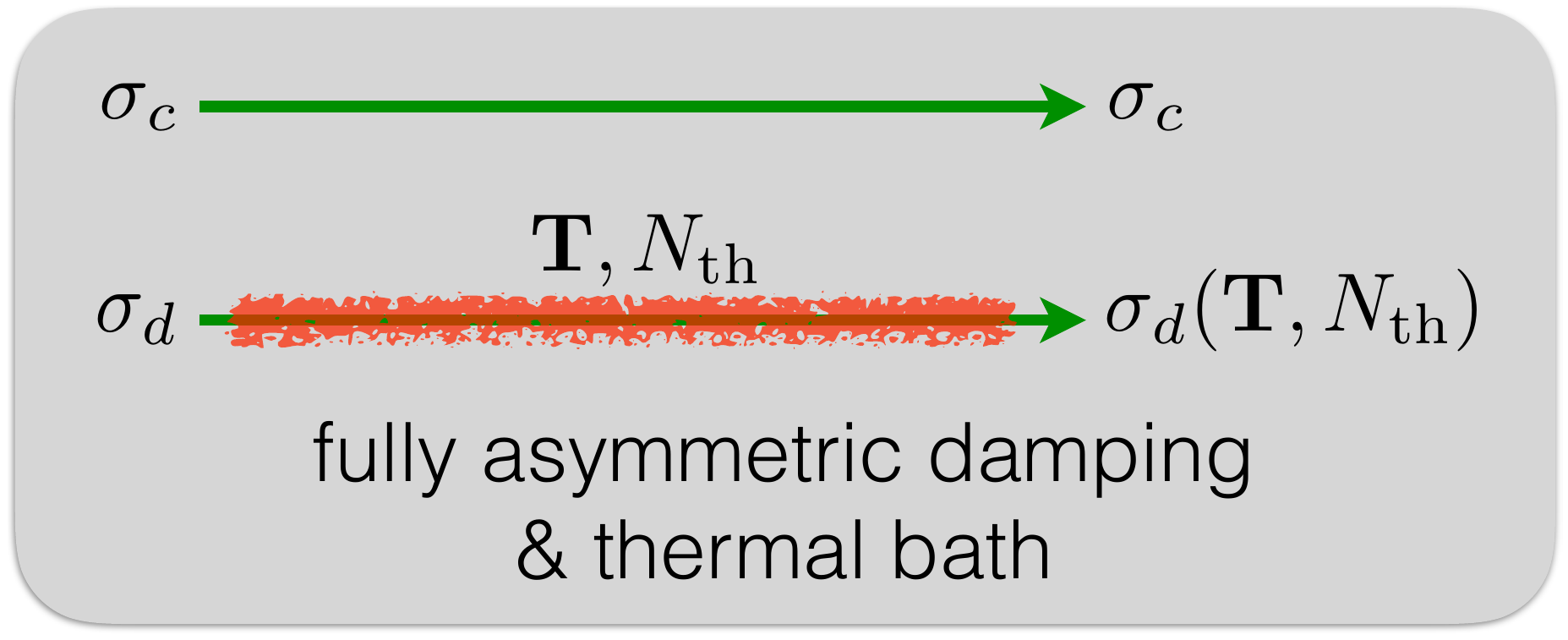}
\end{center}
\caption{
Schematic diagram of the fully asymmetric damping channel. Compared to
the scheme of  Fig.\protect \ref{Fig:asymmetric_scheme2} now mode $d$ is also
coupled to a thermal bath.}
\label{Fig:thermal_asymm_scheme}
\end{figure}

\begin{figure}[ht]
\begin{center}
\includegraphics[width=0.97\columnwidth]{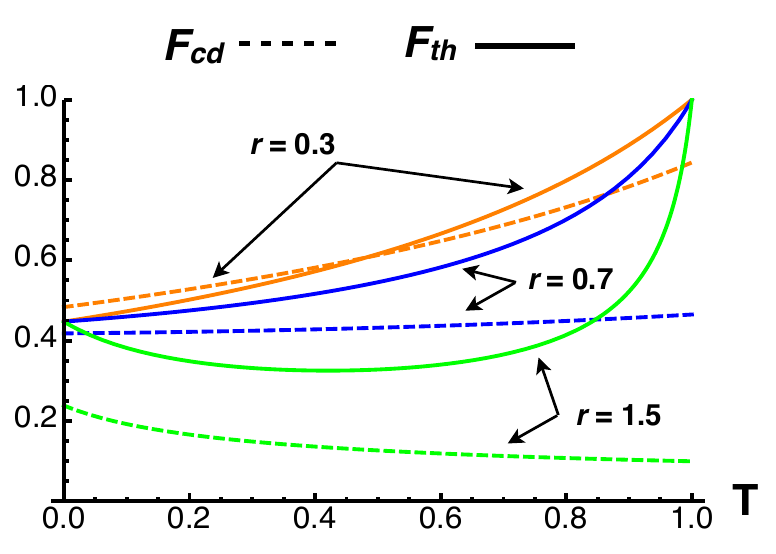}
\end{center}
\caption{Plot of $F_{cd}$ (dashed lines) and the threshold $F_{th}$
(solid lines) as functions of $\mathbf{T}$ for the scheme of
Fig.~\protect \ref{Fig:thermal_asymm_scheme}. We consider three different
values of the initial squeezing $r$. Mode $c$ is prepared in 
a pure squeezed state, whereas mode $d$ travels through a 
damping channel, also coupled to a bath
having $N_{\rm th}=1.0$ average 
thermal photons.} \label{Fig:thermal_asymm_theory}
\end{figure}

The last case we have investigated is the situation where 
asymmetric damping is associated to thermal noise affecting 
one of  the squeezed modes
(see Fig.~\ref{Fig:thermal_asymm_scheme}).
The expected behavior of the fidelities is reported in
Fig.~\ref{Fig:thermal_asymm_theory}.  In this case the left
end of the plot (i.e. for $\mathbf{T}\rightarrow0$) 
corresponds to mixing one squeezed mode to a thermal state 
with an average photon number $N_{\rm th}=1.0$.  
Upon comparing the plot in Fig.~\ref{Fig:thermal_asymm_theory} to 
that of Fig.~\ref{Fig:asymmetric_theory2} one may appreciate the 
effect of adding a thermal bath, i.e. coupling thermal photons to 
a squeezed field.  In this case, both the value of the threshold 
and the fidelity between the two modes decrease with
$\mathbf{T}$. In particular, for small values of the initial 
squeezing (e.g. $r=0.3$) the overall effect of the thermal
contribution is so to make entanglement no more available.
\par
\begin{figure}[h!]
\centerline{\includegraphics[width=0.97\columnwidth]{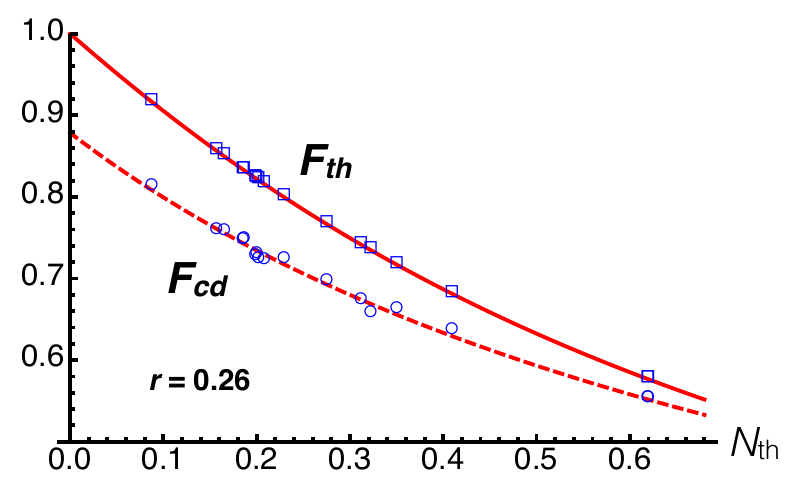}}
\centerline{\includegraphics[width=0.97\columnwidth]{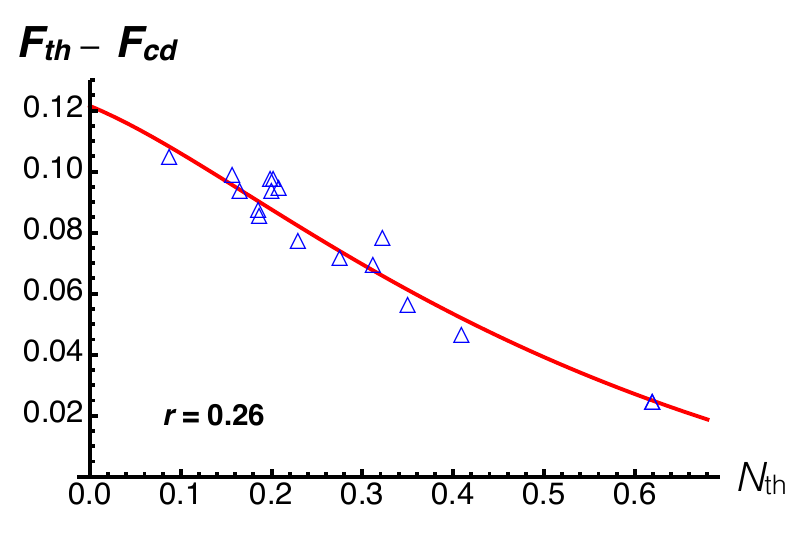}}
\caption{Upper panel: Theoretical threshold fidelity $F_{th}$ 
(solid) and actual fidelity $F_{cd}$ (dashed) together with the
corresponding experimental data (open squares and open circles
respectively) as a function of the average number of bath 
thermal photons $N_{\rm th}$ for the scheme sketched in 
Fig.~\protect\ref{Fig:thermal_asymm_scheme}. Experimental data 
correspond to a set of states with the same value of
the squeezing parameter ($r=0.26$) after the transmission.
Lower panel: Plot of the difference between $F_{th}$ and $F_{cd}$
(those shown in the upper panel) together with the corresponding
experimental data.}
\protect\label{Fig:thermal_asymm_exp}
\end{figure}

We have experimentally investigated the role of thermal photons coupled
to one of the squeezed mode by selecting reconstructed states with the
same measured value of squeezing $r$ (i.e. the residual squeezing parameter
of the state after the transmission) but different thermal contents.
We note that, experimentally, this amounts to considering state coming
from different pure ancestors and undergone to different level of
transmissions. In this way for the same residual squeezing ($r=0.26$
in Fig. \ref{Fig:thermal_asymm_exp})
the effective average number of thermal photon is different. In
the upper panel of Fig.~\ref{Fig:thermal_asymm_exp} we show the fidelity
$F_{cd}$, compared to the corresponding threshold $F_{th}$, as functions
of thermal photons $N_{\rm th}$ effectively coupled into mode $d$.
Actually, these thermal photons are obtained by letting mode $d$ 
propagates in  a lossy channel. In fact, in such a case it is well 
known that a squeezed field
transforms into a squeezed thermal state \cite{JPB2006}.  The results,
reported in Fig.~\ref{Fig:thermal_asymm_exp}, show that the distance to
the threshold decreases as the thermal contribution increases.  
This behavior may be seen more clearly in the lower panel of the same
figure, where we show the difference between the two fidelities.
\section{Conclusions}\label{s:out}
We have experimentally implemented an {\em a priori} certification
scheme based on the fidelity criterion for entanglement generation and
exploited our scheme to assess the effects of losses and noise on the
generation of entanglement by Gaussian states mixing.  In particular, we
have analyzed the effect of signals propagation before the BS and
evaluated threshold values on the transmission coefficient and on the
thermal noise as a function of the parameters of the input signals. We
have considered both symmetric and asymmetric channels with and without
thermal noise.
\par
Our results show that the fidelity criterion represents a reliable tool
for entanglement certification and allows us to accurately take into
account the imperfections of the generation scheme, including
asymmetries and background noise. More generally, our results allow 
one to pre-assess entanglement resources and to optimize  the design of
BS-based schemes for the generation of entanglement for 
continuous variable quantum technology.
\section*{Acknowledgments}
This work has been supported by UniMI through the H2020 Transition 
Grant 15-6-3008000-625. SO and MGAP thanks the late Rodolfo Bonifacio
for his vigorous teaching in the field of quantum optics.

\end{document}